\documentstyle[aps,preprint,eqsecnum,tighten]{revtex}
\newcommand{\ba}{\begin{array}}
\newcommand{\ea}{\end{array}}
\newcommand{\bc}{\begin{center}}
\newcommand{\ec}{\end{center}}
\newcommand{\be}{\begin{equation}}
\newcommand{\ee}{\end{equation}}

\begin{document}

\title{The Level Splitting Distribution in Chaos-assisted Tunneling}

\author{Fran\c cois Leyvraz$^1$\footnote{Permanent address:
Laboratorio de Cuernavaca, Instituto de F{\'\i}sica, Universidad
Nacional Aut\'onoma de M\'exico, Apdo.~postal
20--3643, 01000 M\'exico D.F., Mexico}
and Denis Ullmo$^{1,2}$}

\address{$^1$Division de Physique Th\'eorique, Institut de Physique
Nucl\'eaire, F-91406 Orsay Cedex, France \\
$^2$ AT\&T Bell Laboratories 1D-265, 600 Mountain Avenue, Murray Hill,
New Jersey 07974-0636,USA}
\date{\today}

\maketitle
\begin{abstract}
A compound tunneling mechanism from one integrable region
to another mediated by a delocalized state in an intermediate chaotic
region of phase space was
recently introduced to explain peculiar features
of tunneling in certain two-dimensional systems.
This mechanism is known as chaos-assisted tunneling. We
study its consequences for the distribution of the level splittings
and obtain a general analytical form
for this distribution under the assumption that
chaos assisted tunneling is the only operative mechanism.
We have checked
that the analytical form we obtain agrees with splitting
distributions calculated numerically for a model system
 in which chaos-assisted tunneling is
known to be the dominant mechanism. The distribution depends on two
parameters: The first gives the scale of the splittings and is related
to the magnitude of the classically forbidden processes,
the second gives a measure of the
efficiency of possible barriers to classical transport
which may exist in the chaotic region. If these are
weak, this latter parameter is irrelevant; otherwise
it sets an energy scale at which the splitting distribution
crosses  over from one type of behavior to another. The detailed
form of the crossover is also obtained and found to be in good agreement
with numerical results for models for chaos-assisted tunneling.
\end{abstract}


\section{Introduction}

Under the denomination of ``quantum chaos'', a large body of
theoretical and
experimental work has been, and continues to be, devoted to
the study of the
specific features of a quantum system which can be traced back to the
degree of chaoticity of the underlying classical dynamics
\cite{ozor:book,gutz:book,houches89}.  Among them, questions concerning
tunneling effects receive now an increasing
degree of attention.  Indeed,
although they are often considered as purely quantum, since they correspond
to classically forbidden events,  it appears amply clear
now that tunneling processes are  strongly affected by the nature
of the underlying classical dynamics
\cite{lin,hangi,plata,btu,boh93,tom94}.

This is made most explicit when considering time-dependent
systems as well as systems in more than one dimension,
for which, in contradistinction to the  one-$d$
conservative problems usually considered in
basic quantum mechanics textbooks,
energy conservation does not constrain the motion to be integrable.
A first consequence of having a richer dynamics
is that quasi-degeneracies analogous to those found in the standard
symmetric double well problem
can be observed in systems possessing a discrete symmetry, independently of
whether a potential barrier actually exists or not.  Indeed,
unless the dynamics is entirely ergodic, some classical trajectories
are trapped in $d$-dimensional invariant manifolds (the invariant tori)
inside the $2d$-dimensional phase space.
If there is a discrete symmetry, say parity or time reversal,
in a given system, then any tori which are not themselves
invariant under this symmetry operation will come in symmetrical
pairs. Semiclassical EBK quantization can be applied to these
symmetrical tori, and, since they are entirely identical, this
approximation yields degenerate energy levels. Note that this effect is
distinct from ordinary symmetry induced eigenvalue degeneracy.
Indeed, the effect is only semiclassically correct and tunneling effects
lift this degeneracy. This kind of tunneling has been
dubbed ``dynamical tunneling '' by Davis and Heller \cite{davi81}.

Rephrasing this in terms of wave functions dynamics makes it clearer
why the term ``tunneling'' is proper in spite of the absence of
a barrier.
Indeed, it is possible to construct
what Arnold has termed {\em quasi-modes} \cite{arnold:book}
whenever there exist invariant tori fulfilling the EBK
quantization conditions \cite{maslov:book}.
These are wave-functions semiclassically constructed
on a single torus, and {\em which fulfill the Schr\"odinger
equation up to any order in $\hbar$}.
Here, similarly to what happens when a potential barrier actually
exists (e.g. a one-$d$ conservative system), the actual eigenstates are not
approximated by a single quasi-mode, but rather by a linear combination
of quasi-modes constructed on symmetry related tori.  Therefore, if one
allows a quasi-mode constructed on one of the tori,
to evolve for a very long time,
it will eventually
evolve into its symmetric partner, whereas classically trajectories
remain indefinitely trapped on one torus.

Our interest in higher dimensional systems lies however mainly in
the possibility of considering dynamics of a different nature.
Even for integrable systems, no general theory of tunneling in
multidimensional systems is presently available.  However some theoretical
and numerical studies \cite{wilkinson,crea} clearly demonstrate that
the tunneling mechanism in this case is rather
similar to what is observed for
one-$d$ systems.  In particular the splitting between two quasidegenerate
doublets has a smooth exponential dependence in $\hbar$.

If on the contrary the dynamics is mixed, (as will generically be
the case for
low dimensional Hamiltonian systems), and one is interested in the
tunneling
between two symmetry related invariant tori separated by a significant
chaotic
region, new behavior has been observed, which is
quite different from what one is used to in the
integrable case.{ From} a now growing body of
numerical work, either on one-dimensional time dependent systems
\cite{lin,hangi,plata}, or on two dimensional conservative systems
\cite{btu,boh93}, it has become clear that the presence of chaos
is associated with certain qualitative features, namely:
(i) great enhancement of the average splitting, (ii) extreme
sensitivity to the variation of an external parameter, (iii) strong
dependence of the tunneling properties on what is going on in the
chaotic region separating the two tunneling tori.  A particularly striking
evidence of this last point is given in \cite{boh93}, where it is observed
that reducing drastically the classical transport in the chaotic sea
from the neighborhood of one torus to the one of its symmetric partners
noticeably reduces the tunneling rates.

In \cite{tom94}, it has been demonstrated that a natural interpretation
of this unusual tunneling behavior is obtained if one consider that instead
of the ``direct'' tunneling mechanism characteristic of the integrable
regime, a new tunneling mechanism takes over, in which
the chaotic region plays a predominant role.   In
``direct'' tunneling, splittings are caused by the overlap of the
semiclassical functions constructed on one torus of the symmetrical pair
via the EBK scheme (the  ``quasi-modes''). As discussed above, these
quasi-modes are not eigenstates, but fulfill the Schr\"odinger
equation up to arbitrary order in $\hbar$.  Nevertheless, they can be
connected to other states through  ``tunneling'' matrix elements
which are exponentially small in $\hbar$.
(See section~II.A of \cite{tom94} for an explicit example.)
In the ``chaos-assisted'' regime on the contrary, the picture is that the
particle first tunnels from the integrable region to the chaotic sea and
from there again to the symmetrical region. More precisely, this means that
the tunneling is dominated by the matrix elements between the quasi-modes
and states semiclassically localized in the chaotic region, rather than
by the matrix element connecting directly the two quasi-modes.
This process involves one (or many) intermediate
state and two tunneling processes. The reason why it may nevertheless
dominate
is that, since the chaotic region is much closer in phase
space than the symmetrical integrable region, the tunneling amplitudes
are expected to be usually much larger.
Once in the chaotic region, there is nothing to stop the
particle from reaching the symmetrical integrable region. In this process,
not only do the two semiclassical states play a role, but the
various delocalized chaotic eigenstates which might couple to them
also become relevant. For this reason, the tunneling amplitudes have a
remarkable dependence on $\hbar$: first, they decay exponentially fast with
$\hbar$, reflecting the smooth variation of the tunneling amplitude
from the integrable region to the chaotic sea. Second, however,
and superposed
upon this smooth variation, there is an extremely irregular fluctuation
of the splittings, due to the violent variation in the strength of
the coupling to the nearby-lying chaotic states.
This depends very sensitively on the smallness
of the energy denominator, i.e. on whether or not a chaotic level lies
close to the tunneling doublet. In fact, these fluctuations can
be so strong
as to make any realistic assessment of the $\hbar$ dependence of the
tunneling amplitude impossible.  Note however
that this picture may be oversimplified:
it was observed on the kicked Harper
model by Roncaglia et al. \cite{ronc94}, that direct
tunneling can be
dominant even in the presence of a significant chaotic region.

That the mechanism described above is actually
the one taking place for tunneling in the presence of chaos cannot, at the
present time, be derived from the basic quantum mechanical
law of evolution.  The numerical evidence as well as a far
more careful and detailed discussion of the process are given
in \cite{tom94}.
It is worth stressing,
however, that one of the most important implications of
the chaos-assisted mechanism described above is
that it allows  for a modeling of the splitting distribution
in terms of ensembles of random matrices,
as discussed in more detail in \cite{tom94}.
Therefore not only a qualitative
interpretation of the numerically observed tunneling behavior is obtained,
but also {\em quantitative} theoretical predictions can be made, and
compared to numerically obtained data with a surprisingly good accuracy.

The purpose of this paper being the study of the resulting matrix
ensembles, let us be more specific on what is understood by
``splitting distribution'',
and how the theoretical predictions have been obtained in \cite{tom94}.
As mentioned above, a characteristic feature of tunneling in the presence
of chaos is the extreme sensitivity of such quantities as splittings, to
the variation of an external parameter.  Within the ``chaos-assisted''
interpretation of the tunneling mechanism, this is quite natural since
the splitting of a given doublet may vary by orders of magnitudes depending
on whether a chaotic state is close to the tunneling doublet or not.
Therefore, very small changes of external parameters, which leave
the classical dynamics almost unaltered, may drastically change the
splitting by shifting chaotic levels a distance of a few mean spacings.
Therefore, in any experimental setting the statistical behavior
of the splittings is likely to be of great relevance, even
if one focuses on one single well-defined doublet.
In this case, the physically relevant quantity
will be the distribution of splittings for an ensemble obtained by varying
an external parameter over a range
which is neglible on the classical scale, but still
large on the quantum scale (i.e. the tunneling doublet crosses a lot of
chaotic levels).  By analogy with the random matrix ensembles describing
the spectral fluctuations of classically chaotic systems%
\footnote{These were
originally introduced in Nuclear Physics without any intention to discuss
the nature of the underlying classical dynamics}
(see e.g.~\cite{porter:book,metha:book}), or to their generalization
introduced to describe partly chaotic systems (see section~5 of
\cite{btu}),
one can build random matrix ensembles modeling the tunneling distributions
corresponding to a given phase space structure.

We shall not repeat here in detail the prescription given in \cite{tom94}
for constructing the random matrix ensemble relevant to a given classical
configuration. (A few examples of such ensembles  are given below.)
The following points, however, should be stressed:
First, we are not considering fully ergodic systems
(otherwise there would be no invariant tori); hence the chaotic
part of phase space cannot
a priori be considered as structureless.  A whole
set of partial barriers should quite generally be present, preventing
the motion in the chaotic region from being completely random.
In order to quantify the efficiency of these partial barriers,
additional time scales must be introduced apart from the mean Lyapunov
exponent.   As a consequence,
the random matrix ensembles associated to the chaotic part
of the phase space
cannot be taken as structureless either, as demonstrated in \cite{btu}.
One has to introduce
``transition ensembles'' entirely specified by
a set of ``transport parameters'' $\Lambda_1, \Lambda_2, \cdots$.
These transport parameters
are fixed by the classical dynamics, and are
therefore not adjustable model parameters.
Therefore the parameters which determine the tunneling
distribution are, in the chaos-assisted regime, of two kinds:

\noindent
---(i) the variance $v_t^2$ of the tunneling matrix elements. It describes
the classically forbidden part of the tunneling process. There
exists, at the present time, no theoretical way to evaluate it.

\noindent
---(ii) the set of ``transport parameters'' $\Lambda_1, \Lambda_2, \cdots$,
which are fixed by the classical dynamics inside the chaotic region.

A second thing we would like to stress is that it is not necessary to solve
analytically the thus constructed
random matrix ensembles to obtain a ``theoretical prediction'' for the
splitting distributions.  Indeed the latter are entirely specified
by the random matrix ensemble, and can be obtained concretely by
performing a rather straightforward Monte Carlo calculation:
i.e.~taking at random a large number of matrices with the
distribution specified by the ensemble, diagonalizing them
numerically to obtain the splitting, and construct in this way
a histogram of the splitting distribution.  This was the procedure
used in \cite{tom94} to compare splitting distributions
of doublets of a system of coupled quartic oscillators with those
predicted by the proper random matrix ensemble.

The obvious drawback of using Monte Carlo simulations to produce
the matrix ensembles prediction is that it sheds little light on the
characteristic features of the distribution.  This is made even more
important here because, although all parameters but the tunneling
amplitude
$v_t$ can in principle be computed by studying the classical
motion in the chaotic region, their practical calculation requires
a great deal of effort in the simplest situations, and could turn
out to be impossible for sufficiently complicated classical structures.
Moreover, in experimental realizations, the Hamiltonian governing
the dynamics may not be known in enough detail to allow
fixing the parameters of the ensemble with sufficient confidence.

It is therefore worthwhile to gain some further understanding of the
splitting distribution determined by the
ensembles of random matrices constructed in \cite{tom94}, and
to obtain explicit expressions for these distributions.  As we shall
see, they can in fact be expressed in rather simple form.
Moreover, the resulting distributions have some
very specific general feature, which can be used
as {\em the fingerprint of chaos assisted tunneling} even when
the precise structure of the chaotic motion is unknown.  Because
of the relative complexity of the derivation, we have chosen to organize
this paper as follows.  In section~\ref{sec:result}, we shall give
the final result, without any justifications, and show how well our
analytic findings compare with actual splitting distributions, obtained
in \cite{tom94} for a system of coupled quartic oscillators.
The remaining of the paper will be devoted to the derivation of
this result. Section~\ref{sec:without} will deal with the simpler
case where the chaotic phase space can be taken as structureless.
In section~\ref{sec:with} we shall derive the splitting distributions
in the case where effective partial barriers are present, and discuss
in more details the hypothesis and approximation used in the derivation.
Section~\ref{sec:conc} will be devoted to some concluding remarks.

\section{The splitting distribution}
\label{sec:result}

In the following we introduce some notation.  We are interested
in a system possessing  a discrete symmetry $P$, and for which
tunneling takes place between two quasimodes $\Psi_R^1$ and $\Psi_R^2$
constructed on symmetrical invariant
tori ${\cal T}_1$ and ${\cal T}_2 = P({\cal T}_1)$.
The eigenstates belong to a given symmetry class $+$ or
$-$ depending on whether
they are symmetric or antisymmetric under the action of $P$.
We note $\Psi_R^+$ and $\Psi_R^-$ the symmetric and antisymmetric
combinations of the quasimodes $\Psi_R^1$ and $\Psi_R^2$.
If one neglects the direct coupling between
the quasimodes, $\Psi_R^+$ and $\Psi_R^-$ have the same mean energy
$E_R$.  The ``chaos-assisted'' mechanism proposed in \cite{boh93}
assumes that the tunneling from ${\cal T}_1$ to
${\cal T}_2$ originates from the (exponentially small) coupling
between $\Psi_R^+$  (resp. $\Psi_R^-$) and chaotic states of same
symmetry
$|n,+\rangle$ ($n=1,2,\cdots$) (resp. $|n,-\rangle$) semiclassically
localized
in the chaotic region surrounding the islands of stability containing
${\cal T}_1$ and ${\cal T}_2$.
Therefore, in a basis where the chaotic part of the Hamiltonian is diagonal,
the + and - sectors appear respectively as
	\begin{equation} \label{eq:Hpm}
	H^+ = \left( \begin{array}{cccccc}
	E_r   & v_1^+ & v_2^+ & \cdot & \cdot & \cdot \\
	v_1^+ & E_1^+ &   0   &   0   &       &       \\
	v_2^+ &  0    & E_2^+ &   0   &       &       \\
	\cdot &   0   &  0    & \cdot &       &       \\
	\cdot &       &       &       & \cdot &       \\
	\cdot &       &       &       &       &  \cdot \\
	\end{array} \right)
	\qquad ; \qquad
	H^- = \left( \begin{array}{cccccc}
	E_r   & v_1^- & v_2^- & \cdot & \cdot & \cdot \\
	v_1^- & E_1^- &   0   &   0   &       &       \\
	v_2^- &  0    & E_2^- &   0   &       &       \\
	\cdot &   0   &   0   & \cdot &       &       \\
	\cdot &       &       &       & \cdot &       \\
	\cdot &       &       &       &       &  \cdot \\
	\end{array} \right) \; .
	\end{equation}
After diagonalization of $H^+$ and $H^-$, the regular levels
will be shifted respectively from an amount $\delta^+$ and  $\delta^-$,
giving the splitting
	\begin{equation}
	\delta = |\delta^+ - \delta^-| \; .
	\end{equation}
Eq.~(\ref{eq:Hpm})  merely summarizes the semiclassical picture one has
of the chaos assisted tunneling, but no
random matrix modeling has been introduced yet.
This latter is obtained by assuming that the statistical
properties of the tunneling is correctly reproduced if one model
$H^+$ and $H^-$ by some ensemble of matrices with a specified distribution.

As discussed in detail in \cite{tom94}, the natural choice for
the tunneling
matrix elements $v^\pm_n$ describing
the classically forbidden process is to take them as independent Gaussian
variables with the same variance $v^2_t$, and, in the absence
of any partial barrier, to use for the chaotic levels
$E_1^+, E_2^+, \cdots, E_1^-, E_2^-, \cdots$ the classical ensembles
of Wigner and Dyson which are known to model properly the spectral
statistics of completely chaotic systems
\cite{bgs}. If time reversal invariance symmetry holds, as we shall
asssume in the following, this means that
one should take $E_1^+, E_2^+, \cdots$ and $E_1^-, E_2^-, \cdots$ as two
independent sequences, with a distribution given by the Gaussian Orthogonal
Ensemble (GOE).  Symbolically this ensemble is denoted by
	\begin{equation} \label{eq:without}
	H^+ =  \left( \begin{array}{cc}
	E_R    & \{v\} \\
	\{v\}  &  (GOE)^+ \\
	\end{array} \right)
	\qquad ; \qquad
	H^- =  \left( \begin{array}{cc}
	E_R    & \{v\} \\
	\{v\}  &  (GOE)^- \\
	\end{array} \right) \; ,
	\end{equation}
where the subscripts $+$ and $-$ emphasize the independent nature of the
distribution.  In that case, we shall see in section~\ref{sec:without}
that the splitting distribution $p(\delta)$ is merely a truncated
Cauchy law
	\begin{eqnarray}
	p(\delta) & =  &\frac{4 v_t}{\delta^2 + 4\pi v_t^2}
	\qquad(\delta < v_t) , \label{eq:cauchy} \\
	p(\delta) & = & 0
	\qquad (\delta > v_t)  . \nonumber
	\end{eqnarray}

As demonstrated in \cite{btu}, such a simple statistical modeling of the
chaotic states does not apply any more when structures, such as partial
barriers, prevent classical trajectories to flow freely from one part
of the
chaotic phase space to another.  In such classical configurations (which
are presumably generic in systems where chaos and regularity
coexist), transition ensembles have to be introduced
to obtain a correct statistical description of the chaotic states.
Compared to the case where no barriers are present,
the distribution of chaotic states  will be modified in two ways.
First, each of the parity sequences, taken
separately, will usually not be distributed as a GOE anymore.
However, as stressed
in \cite{tom94}, and as will be made extremely clear in
section~\ref{sec:without}, this has a negligible influence on the
splitting distribution.  More important is that
such barriers may induce strong
correlations between the two parity sequences of chaotic states.
To fix ideas, let us consider a simple example for which a strong partial
barrier separates the chaotic sea into two parts ${\cal R}_1$ and
${\cal R}_2$ which are symmetric images one of each other
under $P$.  In that case, the relevant matrix ensemble can be
symbolically written as \cite{tom94}
	\begin{equation} \label{eq:toy}
	H^\pm =  \left( \begin{array}{cc}
	E_R    & \{v\} \\
	\{v\}  &  (GOE)_S \pm (GOE)_A(\Lambda) \\
	\end{array} \right) \; ,
	\end{equation}
where the variance of the matrix elements of $(GOE)_S$ is chosen such that
it has (in the neighborhood of $E_R$) the same mean spacing $D$ as the chaotic
states,  and the variance
$\sigma^2$ of the matrix elements of $(GOE)_A(\Lambda)$ is fixed by the
transport parameter $\Lambda$ through
	\begin{equation} \label{eq:Ldef}
	\frac{\sigma^2}{D^2} \equiv \Lambda \; .
	\end{equation}
The transport parameter is in turn semiclassically related to the classical
flux $\Phi$ crossing the partial barriers by \cite{tom94}
	\begin{equation} \label{eq:Lflux}
	\Lambda ={1\over4\pi^2}{g\Phi\over(2\pi \hbar)^{d-1}f_1f_2} \; ,
	\end{equation}
where $g=1/2$ is the proportion of states in the corresponding symmetry
class, $f_1=f_2=1/2$ the relative phase space volume  of region
1 and 2 and $d$ the number of freedoms.

 For very ineffective barriers, $\Lambda$ will be much larger than $1$.
$(GOE)_S + (GOE)_A(\Lambda)$ and $(GOE)_S - (GOE)_A(\Lambda)$ will then
be two essentially independent ensembles and  one will recover in that case
the truncated Cauchy law Eq.~(\ref{eq:cauchy}) for the splitting
distribution.  At the opposite extreme, a perfect barrier would give
$\Lambda=0$ (except for
classically forbidden processes), and the $+$ and $-$ spectra will
be strictly identical.  In particular so will be the
displacements $\delta^+$
and $\delta^-$, giving a null splitting.
For small, but finite $\Lambda$, a typical level $E_n^-$ will be usually
found close to its symmetric analog $E_n^+$, though slightly shifted.
In this case, although  the displacements $\delta^+$
and $\delta^-$ are still distributed as Cauchy, they are strongly correlated.
This in turn noticeably affects the splitting $\delta$.

Here it may be usuful to give a qualitative description of the meaning
of $\Lambda$. It can be described as the ratio
$t_H/t_c$ of two timescales, one
of which is purely classical, whereas the other is essentially quantum
mechanical: $t_c$
is the time necessary to cross the barrier, that
is, the typical time that a classical trajectory needs in order to go
from one part of the chaotic sea to the other. The second time scale $t_H$
is the Heisenberg time $\hbar/D$, where $D$ is the average level spacing
in the chaotic sea. Eventually, of course, one expects $\Lambda$
to go to infinity, as the Heisenberg time becomes classically
infinite in the semiclassical limit. Nevertheless
there may well be a very large intermediate region, for
which $\Lambda$ takes very small values. In such cases, there
exists a classical time scale comparable or larger than the Heisenberg
time. This time scale is related to the time necessary to explore
all of the available phase space. In this respect the situation is quite
reminiscent of what happens in localization. The difference, of
course, is that we only have a small number of weakly connected
phase space regions and the long classical time has nothing
to do with diffusion.

Usually, the ensemble describing the classical structure of a system
will be more complicated that the simple one
given in Eq.~(\ref{eq:toy}).  Indeed,
it is probable that transport from one regular island to its symmetric
counterpart is affected through the joint effect of a whole set of
moderately effective barriers rather than by the strong action of a
single one.  Thus one will have to consider much more structured ensembles,
with a transport parameter $\Lambda_n$ associated to each barrier.
Under these circumstances, there is a possibility that localization
effects begin to play a non-negligible role. Should localization
become as effective in limiting
the quantum transport as the classically forbidden processes,
our treatment would become irrelevant. On the other hand  we shall
see in section~\ref{sec:with} that when this
is not the case a noticeable simplification of the problem occurs
because all the information encoded in the transition ensemble
(that is, essentially, the $\Lambda_{n}$'s)
can be summarized in a single parameter $\alpha$, which is
 a weighted average of the variance of the  $(E^+_n - E^-_n)$.

The splitting distribution $p(\delta)$ therefore depends on three
parameters: $(v_t)^2$, the variance of the tunneling matrix elements,
$\alpha^2$ which measures the degree of correlation between the
odd and even levels,  and $D$, the mean spacing of the chaotic levels
to which the $\Psi_R^\pm$ are
connected.  If one consider effective barriers, $\alpha$ is smaller
than $D$.  Moreover, $v_t$ being related to classically forbidden
processes
is usually extremely small, and in particular much smaller than $\alpha$.
We shall therefore assume below $v_t \ll \alpha < D$.
Then, the main result of this paper is that,
for this parameter range, the splitting distribution is given
by:
\begin{itemize}
\item  \underline{ for $ v_t < \delta$}
	\begin{equation} \label{eq:distri:a}
	   p(\delta)=0 \; ,
	\end{equation}
\item \underline{ for $v_t^2/\alpha < \delta < v_t$}
	\begin{equation} \label{eq:distri:b}
	\qquad \qquad  \qquad   p(\delta)=\frac{4 v_t}{\delta^2 + 4\pi v_t^2}
	\qquad  \qquad  \mbox{(Cauchy)} \; ,
	\end{equation}
\item \underline{ for $\delta < v_t^2/\alpha $}
	\begin{equation} \label{eq:distri:c}
	   p(\delta)=2 \mu^{-1} G\left(\frac{\delta}{\mu}\right)
	   \; ,
	\end{equation}
\end{itemize}
where the  function $G$ is the inverse Fourier transform of
$\exp(-\sqrt{|q|})$, namely
	\begin{equation} \label{eq:Gdef}
	   G(x) \equiv \frac{1}{2\pi} \int \exp(iqx) \exp(-\sqrt{q}) dq \; ,
	\end{equation}
	\begin{equation}
	\mbox{and} \qquad
	\mu= \frac{\sqrt{32} \Gamma^2(3/4)}{\pi} \frac{\alpha v_t^2}{D^2} \; .
	\end{equation}
As expected (see section IV.B. of \cite{tom94}), only the smaller
splittings are affected by the transport limitation, the distribution
for larger splittings being unaffected.
The asymptotic behavior of the $\mu^{-1} G(\delta/\mu)$ is given by
	\begin{eqnarray}
	p(0) = 2\mu^{-1} G(0)
	= \frac{1}{\sqrt{2} \Gamma^2(3/4)} \frac{D^2}{\alpha v_t^2}
	& \qquad & \mbox{in $\delta = 0$}
		\qquad \label{eq:G0}  \\
	\mu^{-1} G(\delta/\mu) \simeq  \frac{\Gamma(3/4)}{2^{1/4} \pi}
	\frac{\sqrt{\alpha} v_t}{D} \frac{1}{\delta^{3/2}} \; .
	& \qquad & \mbox{for $\delta \gg \alpha v_t^2/D^2$}
		\qquad \label{eq:Gass}
	\end{eqnarray}
Therefore,  for small enough $\alpha/D$ and $v_t/D$, the distribution
$p(\delta)$ will, in a log-log plot, essentially consist of three straight
lines: (i) a horizontal one (at $p(0)$, as given by Eq.~(\ref{eq:G0})),
for $0 < \delta < \alpha v_t^2 / D^2$. (ii) a line of slope $(-3/2)$
in the range $\alpha v_t^2 / D^2 < \delta < v_t^2/ \alpha$. (iii) a
slope $(-2)$ characteristic of the Cauchy distribution for the range
$v_t^2 /\alpha  < \delta < v_t$, after which the distribution brutally
falls to zero.

Here let us digress shortly to give an intuitive picture of what
is happening. Let us first assume that there are no barriers. Then
chaos-assisted tunneling is essentially a compound process
involving the two symmetrical states and those chaotic states which
lie nearby. Fast tunneling (large splittings) occur only if at least
one of these states actually lies very near to the quasi-degenerate
tunneling state.
This then yields a tunneling process mediated by one single
delocalized chaotic
state. This process has a characteristic $\delta^{-2}$ distribution, as
we shall show later. Here it is essential to realize that the
chaotic state involved,
since it has a well-defined symmetry, will always directly couple
from one torus to its symmetrical partner. On the other hand, if
we have an efficient barrier, the chaotic states also come in quasi
degenerate doublets of opposite parity and of width $\alpha$.
Therefore, moderately fast
processes will be mediated by such a doublet rather than by a single
state. Again, we shall show that this leads to a universal behavior
of $\delta^{-3/2}$ as long as the doublet is identifiable as such, that
is, as long as the two energy denominators contribute roughly equally.
However, for very fast processes, the tunneling will again
be mediated by a single state, namely the one nearest to the tunneling
doublet, and the $\delta^{-2}$ behavior is again obtained. The details
are given by the above formulae, which also show a considerable
amount of information for intermediate cases which cannot be derived
in such a simple fashion.

Before going to the calculation of the above distribution, let us
see how it
compares to actual distributions of splittings obtained numerically for a
system of two coupled quartic oscillators governed by an Hamiltonian
 of the  form
	\begin{equation}
	H({\bf p},{\bf q}) = \frac{p_1^2 + p^2}{2 m}
	 + a (q_1^4/b +b q_2^4 + 2 \lambda q_1^2 q_2^2) \; .
	\end{equation}
Except for their presentation
(we use here a log-log plot instead of a linear versus log binning),
the data
used in Fig.~\ref{fig:qo} are
exactly the same as those used
in Fig.~13 of Ref.~\cite{tom94}, to which we
refer the reader for more precise information on the system investigated.
Here, we shall only say that each set of data has been obtained by
numerically calculating the splittings between regular states constructed
on a given identified pair of symmetrical invariant
tori, for various values
of the coupling $\lambda$.\footnote{In
practice, to increase the statistical
significance of the distribution, data coming from close
lying tori have been
merged.}  The range of variation of $\lambda$ is small on the classical
scale (the classical dynamics remains essentially the same),
but sufficiently
large on the quantum scale that a good statistical significance is reached.

In Fig.~\ref{fig:qo} we display the comparison between the quartic
oscillators data and the predicted form of the distribution
Eqs.~(\ref{eq:distri:a})-(\ref{eq:distri:c}) for two splitting
distributions associated to two different pairs of symmetric invariant
tori.  The agreement is extremely good, especially if one
considers that the distribution extends over more than six decades.
Here a remark is in order.  The parameters $\alpha$ and $v_t$
used for the analytical curves in Fig.~\ref{fig:qo} are actually
tunable parameters.  This was already the case for $v_t$ for the
ensemble introduced in \cite{tom94}, since there is not yet any
semiclassical theory allowing for the calculation of the matrix elements
associated to such classically forbidden processes.
Here however, one has another tunable parameter $\alpha$.  In principle,
this parameter is fixed once the the random matrix ensemble
describing the statistical properties of the chaotic level
is known, which is the case
for this particular system.  In practice however, there is usually no
way to relate $\alpha$ analytically
to, say, the set of transport parameters
$\Lambda_n$'s.  Therefore $\alpha$ eventually plays
the role of a tunable parameter.  It should be borne in mind however
that $\alpha$ and $v_t$ only fix the scale of the distribution, and
in particular the place of the crossover from Cauchy to the $G$-like
behavior, but not its shape.   Therefore, despite the presence of
two tunable parameters, the fact that the splitting distribution
in Fig.~\ref{fig:qo} actually follows the prediction of
Eqs.~(\ref{eq:distri:a})-(\ref{eq:distri:c})
is a very stringent test of the relevance of the whole ``chaos-assisted''
picture.
After this attempt to put the results in perspective, we turn to
their derivation.

\section{The Case Without Barriers}
\label{sec:without}

We now want to get down to computing the distribution of the splittings
in the case in which no barriers are present. Under these circumstances,
 it is
sufficient to compute the distribution of $\delta_+$, and hence $\delta_-$,
since  these splittings are statistically independent.

To start with, one should note that, when considering the matrices $H^+$
and $H^-$ in Eq.~(\ref{eq:Hpm}), the matrix elements $v^\pm_i$ are
associated
with classically forbiden processes, and are thus extremely small.
Therefore, one can compute the displacement $\delta^\pm$ using a first
order perturbation result. One has however here to take special care
with the rare, but important, case, where a chaotic level come extremely
close to $E_R$. This can be done using the exact two
by two diagonalization result for each chaotic eigenstate, and adding up
the contributions.  This gives
	\begin{equation} \label{eq:pert1}
	\delta^\pm = {1\over2} \sum_{i=1}^N (E_R-E^\pm_i)
	\left(1-\sqrt{1+\left({2 v^\pm_i\over E_R-E^\pm_i}\right)^2}\right)
	\end{equation}
(to be understood in the $N \rightarrow \infty$ limit).
In the absence of any values of $E^\pm_i$ too close to $E_R$,
the above equation is equivalent to the usual perturbative result,
	\begin{equation} \label{eq:pert2}
	\delta^\pm 	\approx
	\sum_{i=1}^N {\vert v_i\vert^2\over E_R-E_i^{\pm}} \; ,
	\end{equation}
but the full expression Eq.~(\ref{eq:pert1}) has to be used
to regularize it whenever any of the $(E^\pm - E_R)$ become of the
order of $v_t$.

Although we shall give below a more detailed discussion of that point, the
basic way we are going to use Eq.~(\ref{eq:pert1}) is that the regularized
form of $\delta^\pm(E^\pm_i,v_i^\pm)$ prevents any splitting
from being significantly larger than $v_t$, and that for
$\delta^\pm$ smaller than $v_t$,
Eq.~(\ref{eq:pert2}) can be used safely.
To clarify the discussion, we shall
for the moment replace the $(v^\pm_i)^2$ by their average value
$v_t^2$, and justify below why this does
not change the result. Without loss of generality we
also set $E_R$ equal to zero ($E_R$ is not correlated to
the chaotic spectrum, so it can be used as the origin of the energies).
We shall in addition consider the normalized regular level shift $x$
and energy level $e_1,e_2,\cdots$  (we drop
the superscript $^+$ or $^-$ for the normalized quantities)
	\begin{equation}
	x = {\delta^\pm D \over v_t^2} \qquad ; \qquad
	e_i = \frac{E_i^\pm}{D}  \; .
	\end{equation}

With these manipulations, $p(x)$ is in principle obtained for
$x < D/v_t$ (i.e. $\delta^\pm < v_t$) as the integral
	\begin{equation} \label{eq:integral}
	p(x) = \int \delta \left(x - \sum_{i=1}^N \frac{1}{e_i} \right)
		\, P(e_1,e_2,\cdots,e_N) \, de_1 \, de_2 \cdots de_N \; ,
	\end{equation}
where $P(e_1,\cdots,e_N)$ is is the joint probability of a GOE spectra
with mean density equal to one in the center of the semicircle.
It appears however that,
the correlations of the chaotic states have no influence on $p(x)$,
because the physics here
is determined by the singular nature of the energy denominator which
is not expected to be very sensitive to many-particle correlations.

To demonstrate this point,  let us consider for instance the integral
Eq.~(\ref{eq:integral}) except that we take for the chaotic states
a Poisson distribution, i.e.~that we neglect any correlations between them.
In that case, introducing $\xi_i = N e_i$, one can write
	\begin{equation}
	x = {1\over N} \sum_{i=1}^N \xi_i^{-1} \; ,
	\end{equation}
that is the random variable $x$ is the average of the $\xi_i^{-1}$, where
the $\xi_i$'s are independent variables with density of probability one at
the origin.  Would the distribution of $1/\xi_i$ admit a second moment
(i.e.~$\langle \xi^{-2} \rangle < \infty$), the usual central limit theorem
would yield a Gaussian distribution for $p(x)$.  Here, however,
the situation is quite different since this variance actually diverges.
The distribution $p_0(y)$ of the $y_i=\xi_i^{-1}$ behaves as $y^{-2}$
for $y \gg 1$, whatever the initial distribution of the $\xi_i$,
as long as that distribution is equal to one for $\xi$ equal to zero%
\footnote{Therefore, the result will not be affected by a secular change of
the mean density of states away from the origin.}
 From this follows through standard probabilistic arguments \cite{Feller}
that $x$ has a non-singular limiting distribution, namely the Cauchy
law
	\begin{equation} \label{eq:cauchy1}
	p(x)={1\over\pi^2+x^2} \; .
	\end{equation}
Informally, this result can be  obtained as follows. If the chaotic states
are distributed independently, Eq.~(\ref{eq:integral}) reads
	\begin{equation} \label{eq:int2}
	p(x) = \int_{-\infty}^\infty
	 \prod_{i=1}^N dy_i\, p_0(y_i)
	\delta\left(x-{1\over N}\sum_{i=1}^N y_i\right) \; .
	\end{equation}
If we now take the Fourier transform of Eq.~(\ref{eq:int2}),
the result factorizes and one obtains
	\begin{equation}
	\hat p(q) \equiv \int_{-\infty}^\infty p(x) e^{iqx}dx
	= \hat p_0(q/N)^N \; ,
	\end{equation}
where $\hat p_0(q)$ is the Fourier transform of $p_0(y)$. The large-$y$
behavior of the latter leads to a singularity of $\hat p_0(q)$, which, by
reason  of the symmetry of $p_0(y)$, must be located at the origin.
Further, this singularity is of the type of a discontinuous derivative.
For any symmetric function  $f(y)$ with the same large $y$ behavior as
$p_0(y)$, $f(y)-p_0(y)$ decreases more rapidly that $y^{-2}$, which implies
that
$\hat f(q) - \hat p_0(q)$ has a continuous derivative.  The jump in $q=0$
of the derivative of $\hat p_0(q)$ must therefore be the same as for
$f(y) = (1+x^2)^{-1}$, the Fourier transform of which is  $\exp(-\pi|q|)$.
Noting moreover that,
because of the normalization, $\hat p_0(0) = 1$, one has
	\begin{equation}
	\hat p_0(q)=1-\pi\vert q\vert+o(q)\qquad(q\ll1) \; ,
	\end{equation}
and therefore
	\begin{equation}
	\hat p(q) = \lim_{N \to \infty} \left( 1 - \frac{\pi \vert
        q\vert}{N} \right)^N = \exp(-\pi \vert q\vert),
	\end{equation}
from which the result follows immediately by inverse Fourier
transformation.

At the opposite extreme, one can consider the most rigid spectrum, and
see what happens if the chaotic states are distributed as a picket fence.
In that case, $p(x)$ can be written as
	\begin{equation}
	p(x) = \int_{-1/2}^{+1/2} de \,
	\delta \left(x -  \sum_{n=-\infty}^{+\infty}
	\frac{-1}{n+e}\right) \;
	\end{equation}
which, using the equality \cite{gradstein}
	 \[ {\rm cotg}(\pi x) = \frac{1}{x} + \frac{2x}{\pi}
	\sum_{k=1}^\infty
	\frac{1}{x^2-k^2} \]
readily gives
	\begin{equation}
	p(x) = \int_{-1/2}^{+1/2} dE \,
	\delta \left(x -  \pi {\rm cotg}(\pi E) \right)
	= \frac{1}{x^2 +  \pi^2}  \; ,
	\end{equation}
that is the very same Cauchy distribution as for the Poissonian case.
There is no doubt that if the two extremes of totally uncorrelated  and
completely correlated spectra give the same result, the correlation
between
chaotic states play little or no role.  Therefore, as demonstrated
in Fig.~\ref{fig:without}, the splittings are
also Cauchy-distributed
when the chaotic states are GOE distributed.  Indeed, this
result can actually be shown using supersymmetric techniques
\cite{vonOppen} and also turns out to follow from results
on $S$-matrix ensembles for the one-channel case \cite{lopez}
under quite general conditions for both the $v_i$ and the
energies $E_i$. In fact, it turns out that the sums involved
in computing the $K$-matrix in a one channel system  are exactly
of the type we are interested in and their distribution can be
found exactly under the assumption that the ensemble is ergodic
and analytic in the energy. For details see \cite{lopez} and
\cite{mello}.

Before turning to the more difficult case of problems where transport
limitations play a role in the tunneling mechanism, let us come back to
a couple of points not treated in the above discussion.  Since we have seen
that correlations between chaotic states are of little importance, we
shall discuss these points
under the assumption that there are no such correlations.
The first concerns the fact that the
tunneling matrix elements are randomly distributed following
a Gaussian law, instead of being constant as assumed in the above
discussion.  However, it can easily be checked that, in the Poissonian
case, this simply amounts to performing first the integral over the
tunneling matrix
elements distribution.  The second point concerns the need to use
the regularized form Eq.~(\ref{eq:pert1}) instead of its nondegenerate
approximation Eq.~(\ref{eq:pert2}). Let us now see the effect
of using this more correct
formula which takes quasidegeneracies fully into account by treating the
corresponding $2\times2$ matrix exactly. In this case the relevant function
of $\xi_i$ is equal to
	\begin{equation}
	y_i = \xi_i \left( (C_N)^2 - \sqrt{ (C_N)^4 +
		\left({2v_i\over\xi_i}\right)^2} \right)
	\qquad ; \qquad C_N = \frac{D^2 N^2}{v_t^2}
	\end{equation}
which is equal to $\xi_i^{-1}$ for $\xi_i\gg C_N^{-1}$ but
saturates to a value of $1/v_t$ for smaller
values of $\xi_i$. This implies that $\hat p_0(q)$
has a singularity of the type described above only for $q$ less than
$v_t/(N D)$. This in turn involves a departure of $\hat p(q)$
from pure exponential behavior when $q$ becomes of  the order of
$v_t/D$, and hence for normalized splittings of the  order of
$D/v_t$, which in unnormalized units correspond to splittings of order
$v_t$.   We recover in this way the intuitive picture discussed above,
namely that the basic role of the regularization is
to prevent having splittings of size larger than the root
mean square deviation of the $v_i$, whereas the distribution
for smaller values remains unaffected.
For the distribution $p(x)$ displayed in Fig.~\ref{fig:without},
the effect of this regularization cannot be observed because it only
affects the range $x \geq v_t^{-1} = 10^4$ which is not covered
in this linear scale.  It is however clearly seen in Fig.~\ref{fig:qo},
as well as in the Figs.~\ref{fig:MCtoy} and \ref{fig:MCqo} of the
following section.

\section{The Case of Efficient Barriers}
\label{sec:with}

Let us now consider the more complicated case of systems for which
transport limitation induces strong correlation between the symmetry
 classes.
As we have emphasised in the previous section, the correlation between
chaotic states inside a symmetry class has little or no influence on the
distribution of the shifts, $\delta^+$ and $\delta^-$, of the regular
level, due to their coupling with the chaotic states.
Therefore, even in the case where there exist efficient
barriers to transport in the chaotic
region, the shifts $\delta_\pm$ should be
still distributed according to the Cauchy distribution
derived  in the preceding section. In fact, the main effect of such
barriers is to induce strong correlations between the
$E_i^+$ and the $E_i^-$, which only affect the distribution of the
splittings themselves $\delta = |\delta^+-\delta^-|$.
Another consequence of the presence of partial barriers which may also
influence the distribution of splittings is that it may yield some
inhomogeneity of the variance of the tunneling matrix elements, as
well as of the correlation between chaotic states.
We shall come back to this point at the end of the section when comparing
our findings with exact results calculated numerically using Monte-Carlo
techniques.

\subsection{Derivation of
Eqs.~(\protect{\ref{eq:distri:a}})-(\protect{\ref{eq:distri:c}}) }
\label{sec:DerOfEqs}

To lighten somewhat the notation we use in this subsection scaled
energies $e^\pm_i = E_i^\pm / D$ (for which the mean density of states
around the regular level is therefore one), and note $\bar v_t = v_t/D$
and $\bar \alpha = \alpha/D$.
We  use the following modeling of our problem.
First, we shall ignore any correlation between chaotic states inside each
symmetry class and merely require that the mean density of chaotic
states be equal to $D$ arround $E_R=0$.  To normalize the number of states
to $N$, we choose the variables $e^+_i$ and
$e^-_i$ both distributed according to a Gaussian law, which we take to
be $N^{-1} e^{-\pi (e^\pm_i/N)^2}$.
In this way, the density at zero is $1/N$ for
each level (and thus the total density is $1$).   Since we shall consider
the $N \rightarrow \infty$ limit, this can be though of as a flat
distribution on the scale
of a mean level spacing, the Gaussian form being just
introduced to normalize in a proper way the distribution.
Again, we take the chaotic states to be identically distributed
random variables, the $e^+_i$ are independent of each
other for different values of $i$.  The correlations between the two parity
sequences of chaotic states must, however, be implemented, which we shall
do assuming that
the $(e^+_i - e^-_i)$ have a Gaussian distribution characterized
by its width $\bar \alpha$.  As discussed in section~\ref{sec:result}
$\bar \alpha$ is related to a characteristic time necessary for a classical
trajectory to travel from one regular island
to its symmetric counterpart.
To justify  this construction, let us consider for instance the
ensemble of Eq.~(\ref{eq:toy}) introduced in section~\ref{sec:result}.
In this case, it was seen that  the Hamiltonians of the two
symmetry sectors
are related to one  another by adding a GOE matrix, the
off-diagonal elements  of
which have variance $(2\sigma)^2$, related through
Eqs.~(\ref{eq:Ldef})-(\ref{eq:Lflux}) to the classical flux $\Phi$
crossing the partial barrier.   As is well-known, the resulting spectrum
is formally the same as the result of letting the levels
moving according to an interacting Brownian motion
during a time $\Lambda=\sigma^2/D^2$, as described by Dyson \cite{dyson}.
 However, for short times (i.e.~small $\Lambda$),
it is generally accepted that an interacting diffusion process
can be replaced
by a free one \cite{free_diff}, which here means that the $(E^+_i-E^-_i)$
follow a Gaussian distribution, of width $\alpha=\sqrt{2} (2\sigma)$.
This also follows more specifically from the results shown
in \cite{self_cons}. There it was shown that a randomly distributed
sequence diffusing over a short time $\Lambda$ only acquires
correlations on a scale of $\sqrt\Lambda$.
(Note that for $\alpha^2 \gg D^2$, this modeling becomes essentially
meaningless.   For instance it is not possible anymore to specify
unambiguously what $E_{i'}^-$  is to be associated with a given $E_{i}^+$
when considering the variance of their difference.  This is not
of great importance, though, since the final
result shows that in this case the truncated Cauchy
behavior described in section~\ref{sec:without} is recovered anyhow.)
Thus  the joint
probability distribution function of the $e^+_i$ and the $e^-_i$ is
	\begin{equation} \label{eq:joint}
	P(e^+_i,e^-_i) = \left({1 \over \sqrt{2\pi}
        \bar \alpha N}\right)^N
	\prod_{i=1}^N \exp\left (-\pi \left(\frac{e^+_i}{N}\right)^2 +
	{1\over 2\bar\alpha^2}(e^+_i-e^-_i)^2\right) \; .
	\end{equation}
Note the asymmetric treatment of $e^+_i$ and $e^-_i$. This
simply means that, because of the very strong correlation between
the two sets of eigenvalues, the large-scale distribution of one
set entirely determines that of the other.

For more structured ensembles than the one of Eq.~(\ref{eq:toy}), the
actual diffusion process, while ``turning on'' the transport parameters
$\Lambda_n$'s from zero to their actual value, may be noticeably more
complicated.  Nevertheless, because the final splitting results from
the average effect of the coupling with a large number of chaotic states,
it is natural to assume that a kind of central limit theorem is involved
and that the form Eq.~(\ref{eq:joint}) can also be used in practice
(we shall discuss this question in greater detail in the next subsection).
As mentioned in section~\ref{sec:result} however, there  will not be
anymore a
simple relationship between $\alpha$ and the transport parameters
of these ensembles.

Now the problem is to compute the splitting distribution. We
shall disregard in the following the complications created
by the inclusion of the complete expression
Eq.~(\ref{eq:pert1}) for the shifts, since
this only causes a cutoff at  values of $\delta$
equal to $v_t$, as was already discussed
in the previous Section.
As a further simplification we shall take for a moment the
tunneling matrix elements as being constant (equal to $v_t$) and shall
come back later to the (slight) modifications to the result due to
averaging over their Gaussian distribution.
Introducing the scaled variable $X = (\delta^+ - \delta^-)/D$,
($\delta / D = |X|$) the distribution of $X$ is obtained as
		\begin{equation} \label{eq:distri_def}
	p(X) = \int \prod_{i=1}^N de^+_i\,de^-_i\,P(e^+_i,e^-_i) \,
	\delta\left(X  - \bar v_t^2 \sum_{i=1}^N
	(\frac{1}{e^+_i}-\frac{1}{e^-_i})
	\right) \; .
	\end{equation}
Again, this integration can be factorized by introducing the Fourier
transform $F(q)$ of $p(X)$, and
everything can be reduced to quadratures. The details are a trifle tedious
and are therefore relegated to Appendix A. The final result is
	\begin{eqnarray}
	p(X) & = & {1\over2\pi} \int_{-\infty}^\infty F(q)  e^{-iqx} dx
	                       \label{eq:Fdef} \\
	F(q) & = &
	    \exp\left(-{\bar \alpha \over\sqrt{2 \pi} } \Psi(\tilde q)
		 \right)
	    \qquad ; \qquad \tilde q = \frac{\bar v_t^2}{\bar \alpha} q
	     \label{eq:from_app} \; .
	\end{eqnarray}
Here $\Psi(\tilde q)$ is given by the expression
	\begin{eqnarray}
	\Psi(\tilde q) & = & \int_{-\infty}^\infty dy\,
	   \Phi\left({\sqrt{8} \tilde q\over|1-y^2|}\right) \; ,
	   \label{eq:psi} \\
	\Phi(z) & = & 2\int_0^\infty{dt\over t^3}(1-\cos zt)e^{-1/t^2} \; .
	\end{eqnarray}
For ease of reference we give the following integrals, which
are derived for completeness in Appendix B:
	\begin{eqnarray}
	  \int_{-\infty}^\infty\Phi(1/y)dy &=& \pi^{3/2}
	    \label{eq:Bdeux}\\
	  \int_{-\infty}^\infty\Phi(1/y^2)dy &=&\sqrt{2\pi}\,\Gamma({3/4})
	   \label{eq:Bun}
	\end{eqnarray}
An asymptotic study of the function $\Phi(z)$ for $z \gg 1$ and $z \ll 1$
yields
	\begin{eqnarray}
	\Phi(z) & = & -z^2 \ln z +O(z^2) \qquad (z \ll 1) \cr
	        & = & + 1 + O(z^{-1}) \qquad \qquad (z \gg 1) \; ,
	\end{eqnarray}
(the prefactors given here are actually correct, but
we shall not need them, and the order of magnitude is easy to obtain).
Therefore, $\Psi(\tilde q)$ basically gives a measure of the domain of $y$
such that  $\tilde q/(1-y^2)$ is larger than one. For $\tilde q \gg 1$,
this is obviously of order $\sqrt{\tilde q}$.  It can moreover be evaluated
more precisely by noting that in this range of $\tilde q$,
$ \Phi\left({\sqrt{8} \tilde q / |1-y^2|}\right)
\simeq \Phi\left({\sqrt{8} \tilde q / y^2}\right)$. This is the case,
because in the range
of $y$ where $|1-y^2|^{-1} \not \simeq y^{-2}$, i.e. for $y$ not large,
both $ \tilde q / |1-y^2|$ and $ \tilde q / y^2$ are much larger than one
if $\tilde q$ is,
and therefore $\Phi$ saturates to its asymptotic value $1$ anyhow.
One therefore finds for $\tilde q \gg 1$
	\begin{equation}  \label{eq:qgg1}
	\Psi(\tilde q) \approx
	\int_{-\infty}^\infty dy\,
	 \Phi\left({\sqrt{8} \tilde q\over y^2}\right) =
	2^{3/4} \Gamma(3/4)  \sqrt{2 \pi \vert \tilde q\vert } \; .
	\end{equation}
and therefore
	\begin{equation}  \label{eq:F:qgg1}
	F (q) \approx F_\infty(q) = \exp \left(-2^{3/4} \Gamma(3/4)
	 \sqrt{\bar \alpha \bar v_t^2\vert q\vert } \right)
	\qquad \qquad \mbox{for }  q \gg \frac{\bar \alpha}{\bar v_t^2}\; .
	\end{equation}
For $\tilde q \ll 1$ on the other hand, $\tilde q /(1-y^2)$ is large
only in the neigborhood of $y=1$.
 From this follows that one can restrict oneself to the range of
integration $y \sim \pm 1$.  One obtains
	\begin{eqnarray}
	\Psi(\tilde q) & \approx & 2 \int_{1 - \tilde q}^{1+\tilde q} dy\,
             \Phi \left(\sqrt 8  \tilde q \over (1-y) (1+ y) \right)
				\nonumber \cr
	               & \approx & 2 \int_{-\infty}^\infty dt\,
	 \Phi \left(\sqrt 2 \tilde q \over t\right) = (2\pi)^{3/2} \tilde q
	     \label{eq:qll1} \; ,
	\end{eqnarray}
and therefore
	\begin{equation}  \label{eq:F:qll1}
	F (q) \approx F_0(q) =
	 \exp \left(-2 \pi \bar v_t^2 \vert q\vert \right)
	\qquad \qquad \mbox{for }
	q \ll \frac{\bar \alpha}{\bar v_t^2} \\; .
	\end{equation}
It remains to perform the inverse Fourier transform
Eq.~(\ref{eq:from_app}),
and to deduce the asymptotic behavior of $p(X)$ from the one of
$F (q)$.  For large $X$, $p(X)$ is dominated by the singularities
of its Fourier transform, which here means the derivative discontinuity at
the origin.  Therefore for $\bar v_t>X \gg \bar v_t^2 /\bar \alpha$
one can use in Eq.~(\ref{eq:from_app}) the asymptotic
$ q \ll {\bar \alpha}/{\bar v_t^2}$ approximation $F_0(q)$ of $F(q)$.
Applying the inverse Fourier transformation yields an almost perfect
Cauchy distribution of the form
	\begin{equation} \label{eq:cauchy2}
	p(X) =  \frac{2 \bar v_t}{X^2 + 4\pi^2 \bar v_t^2} \; .
	\end{equation}
The above result amounts to adding
independently the variable $\delta^+$ and
$\delta^-$, distributed as given by Eq.~(\ref{eq:cauchy1}), and to neglect
the correlations between the two symmetry classes.
As mentioned in \cite{tom94}, this is indeed quite natural since
splittings larger than $v_t^2/\bar\alpha$ are due to chaotic levels coming
closer than a distance $\bar\alpha$ of the regular level. Since
$\bar\alpha$ can be viewed
as the scale on which chaotic levels are correlated,
chaotic states contributing to $p(X)$ for $X \gg \bar v_t^2 / \bar \alpha$
can therefore be considered as essentially
decorrelated from their symmetric
counterparts. In the language of section 2, the splittings we are
looking at here are so large that the tunneling is always mediated by
a single state.

Let consider now the range $X \ll \bar v_t^2/\bar \alpha$, for which
the splitting distribution is affected by correlation between symmetry
classes.  In that case, the term $\exp(-iqX)$ in the inverse Fourier
transformation Eq.~(\ref{eq:Fdef}) is essentially constant in all the range
$q \leq \alpha/v_t^2$ where $F(q)$  differs from its
asymptotic behavior $F_\infty(q)$. Noting that
	\begin{equation}
	{1\over2\pi} \left(\int_{-\infty}^\infty
	F_\infty(q) \exp(-iqx)dq \right) = \lambda^{-1} G(X/\lambda)
	\qquad ; \qquad \lambda=\sqrt 8 \Gamma^2(3/4)
	\bar \alpha \bar v_t^2 \; ,
	\end{equation}
where
$G(x)$ is the inverse Fourier transform of $\exp(-\sqrt q)$ as defined in
Eq.~(\ref{eq:Gdef}). One therefore has
	\begin{equation}
	p(X) = \lambda^{-1} G(X/\lambda) \; + \; K
	\end{equation}
where $K$ is the constant
	\begin{equation}
	K \equiv \frac{1}{2\pi}
	\int_{-\infty}^\infty  dq \left(F(q)-F_\infty(q) \right)
	\end{equation}
For small $\bar \alpha$ however, $K$ is of order
$\bar \alpha^2/\bar v_t^2$, when $\lambda^{-1} G(X/\lambda)$ range
from order $1/(\bar \alpha \bar v_t^2)$ at $X=0$ to
$\bar \alpha^2/\bar v_t^2$ at its lowest value, i.e. at the
crossover $X \sim \bar v_t^2 / \bar \alpha$
between the Cauchy-like and G-like behavior.  Therefore it can usually be
neglected, although in some special circumstances it shows up as a small
plateau between these two regimes; we shall disregard it from now on.
Then, the large-$X$ behavior of  $\lambda^{-1} G(X/\lambda)$ is dominated
 by the $\sqrt q$ singularity at the origin of $F_\infty(q)$,
so that it goes as $X^{-3/2}$ as $X\to\infty$.
This is in fact the hallmark of the $X \ll \bar v_t^2 / \bar \alpha$
regime we are discussing.

Finally, one has to take into account the fact
that the tunneling matrix elements are not constant, but randomly
distributed.
As can be seen in the derivation of Eq.~(\ref{eq:from_app}) (see the remark
below Eq.~(\ref{eq:eq_app})) this merely amounts to replacing the
expression
for $\Psi(q)$ given in eq.~(\ref{eq:psi}) by the function $\bar\Psi(q)$
defined as follows:
	\begin{equation}
	\bar\Psi(q)=\left\langle\Psi\left({v^2q/\alpha}\right)
	\right\rangle_v,
	\end{equation}
where the brackets denote averaging over the $v$'s, which we take
to have a Gaussian distribution with variance $v_t^2$. This new function
is of course much more complicated than the original one,
but its asymptotic behavior for small or large values of $q$ is readily
obtained. Indeed, for $q\ll1$, $\Psi(q)$ is proportional to $\vert
q\vert$, so that
the average is obtained by replacing $v$ by $v_t$. On the other hand, for
$q\gg1$, $\Psi(q)$ is proportional to $\sqrt{\vert q\vert}$,
so that its average over a Gaussian distribution is obtained by
replacing $v$ by $\sqrt{2/ \pi}v_t$.
Using these facts together with the above estimates for the behavior
of $p(X)$ one finally obtains the result stated in
Eqs.~(\ref{eq:distri:a}--\ref{eq:distri:c}) in
section~\ref{sec:result}. Some trivial differences: There we consider
splittings as being always positive, whereas in the above computation
we treated positive and negative splittings separately: this
introduces a new factor of two. Further, we have made the dependence
on $D$ explicit, which in particular means replacing $p(X)$ by
$Dp(X/D)$ as well as replacing $\bar\alpha$ and $\bar{v_t}$ by their
original expressions.

\subsection{More Structured Ensembles}

For the simple
ensemble Eq.~(\ref{eq:toy}), the two main assumptions we made
to replace the exact distribution by Eq.~(\ref{eq:distri_def}),
namely to neglect correlation of chaotic states among a given
symmetry class and to replace the
interacting brownian diffusion by a free one for small
$\bar \alpha$ are really under control.  Indeed, section~\ref{sec:without}
 gives full justification of the first assumption, and the second can be
seen as a
simple consequence of standard perturbation theory.  And actually,
as shown in Fig.~\ref{fig:MCtoy}, one can see that our analytical
findings perfectly agree wit an ``exact'' Monte-Carlo evaluation of the
splitting distribution generated by the ensemble Eq.~(\ref{eq:toy}).
We stress that in this very simple case the parameter $\alpha$
that we are using is (for small $\alpha$) simply related to the variance
$\sigma^2$ of the non-diagonal matrix elements of $(GOE)_A$
(indeed $\alpha^2 = 2 (2\sigma)^2$), and that therefore there are no
adjustable parameters in this comparison.

More structured ensembles deserve however some further discussion.
Let consider
for instance the ensemble relevant to the quartic  oscillator system
used as illustration in section~\ref{sec:result}. Symbolically,
this ensemble can be written as \cite{tom94}
	\begin{equation} \label{eq:qo_struct}
	H^\pm_{\rm qo} =  \left( \begin{array}{cccc}
  E_R  & \{v\}                  &           0         &        0        \\
\{v\}  &  (GOE)_1     & (GOE)^\pm(\Lambda_{12}) & (GOE)^\pm(\Lambda_{13})\\
 0     & (GOE)^\pm(\Lambda_{12})&        (GOE)^\pm_2  &         0        \\
 0     & (GOE)^\pm(\Lambda_{13})&           0         &      (GOE)^\pm_3 \\
	\end{array} \right) \; ,
	\end{equation}
where the subscript $\pm$ again indicates ensemble which are independent
in the $+$ and $-$ symmetry class.
Noting $D_{\rm tot}$ the total density of states (in a given
symmetry class),
$(GOE)_i$ stands for a Gaussian Orthogonal Ensemble such that
the mean level
density in the center of the semicircle is $f_i D_{\rm tot}$, and
$(GOE)^\pm(\Lambda_{jk})$ represent Gaussian distributed independent
matrix elements of variance $\sigma_{jk}^2 = \Lambda_{jk} D^2_{\rm tot}$.
(For the configuration of the quartic oscillators corresponding to
Fig.~\ref{fig:qo}, one has $f_1=0.5, \;  f_2=0.2, \; f_3=0.3$, and
$\Lambda_{12} = 0.14, \; \Lambda_{13}=0.11$.)
For such complicated ensembles the two assumptions
concerning the irrelevance of intra-class correlations and essentially
Gaussian distribution of the $(E^+_i-E^-_i)$ are presumably equally
well
fulfilled here as in the simple case of Eq.~(\ref{eq:toy}).  What is
lost however is the uniformity of the distribution of the tunneling matrix
elements and of the variance of the $(E^+_i-E^-_i)$.  Indeed, in the above
example, a diagonalization of the chaotic part of the Hamiltonian is
going to transfer some tunneling matrix elements from the block connecting
$E_R$ to $(GOE)_1$ to the ones connecting $(GOE)_2$ and $(GOE)_3$.  One may
end in this way with three different scales for the variance of
the tunneling matrix elements as well as for the parameter $\alpha$
(one for each (GOE) block).

More generally, the typical situation will be that a [possibly large]
number of $(GOE)$ blocks, $(GOE)_1$, $(GOE)_2$, ..., $(GOE)_K$ are involved
in the tunneling process.  After diagonalization
of the chaotic part of the Hamiltonian, both the variance of the
tunneling matrix elements, and the degree of correlation between
symmetry classes, will be block dependent.  Each block $(GOE)_k$
($k=1,\ldots,K$) would have then to be characterized by a tunneling
parameter $v_k$ and a transport parameter $\alpha_k$ ($\alpha_k$
and $v_k$ highly correlated), in addition to
its dimension $N_k = f_k N$.  Let us introduce the notation
	\begin{equation}
	I(\alpha,v_t;q) = -{\alpha \over D\sqrt{2 \pi} } \Psi(
	v_t^2q/\alpha D)
	\; , \end{equation}
where $\Psi(\tilde q)$ is defined by Eqs.~(\ref{eq:from_app}).
A straightforward modification%
\footnote{
In Eq.~(\ref{eq:eq_app}), $\lim_{N \to \infty} (1-I(q)/N)^N$
has to be replaced by
\[ \lim_{N \to \infty} \left[\prod_{k=1}^K
(1-I(\alpha_k,v_k;q)/N)^{f_k N} \right]. \]
}
 of the derivation
of Eq.~(\ref{eq:from_app}) gives that taking into account the block
dependence of $\alpha_k$ and $v_k$ merely amounts to replacing this
equation (i.e. $F(q)= \exp(-I(\alpha,v_t;q))$) by
	\begin{equation}
	F(q) = \exp \left(- \sum_{k=1}^K f_k I(\alpha_k,v_k;q) \right)
	\; . \end{equation}
Inspection of Eqs.~(\ref{eq:F:qgg1}) and (\ref{eq:F:qll1}) then shows
that they remain valid provided $\alpha$ and $v_t$ are defined now
as
	\begin{eqnarray}
	v_t^2 & \equiv & \sum_k f_k v_k^2 \label{eq:equi_v} \\
	\alpha  & \equiv & \frac{1}{v_t^2} \left(\sum_k f_k
		\alpha_k^{1/2} v_k\right)^2 \; .  \label{eq:equi_a}
	\end{eqnarray}
Multiplying Eq.~(\ref{eq:equi_v}) by $N$,
$N v_t^2$ appears as the [average] square norm  of the projection of the
quasimode $\Psi^\pm_R$ on the chaotic space.  It is therefore
independent of the chaotic phase space
structure.  This, for instance, allows to compute $v_t$ from
the variance of the tunneling matrix elements {\em before}
diagonalization of the chaotic part of the Hamiltonian.
The parameter $\alpha$ and $v_t$ have moreover
a certain number of intuitively clear properties:
If all $v_k$ are multiplied by a constant factor, the effective
tunneling element $v_t$ is multiplied by the same factor, whereas
$\alpha$ is unaffected.  Further, if all $v_k$ are identical, then
the effective tunneling element is the same. On the other hand, the
same is not true of $\alpha$: If all $\alpha_k$ and all $v_k$ are
taken to be equal, the effective efficiency of the classical barrier
now depends on the number of different components of phase space
through which tunneling can take place. Further, we see that any components
with negligible values of $v_k$ will contribute negligibly
both to $v_t$ and $\alpha$. Thus we can identify a given part of
phase space through which tunneling actually occurs and limit
ourselves to it.

With the definitions Eqs.~(\ref{eq:equi_v}) and (\ref{eq:equi_a}) of
$v_t$ and $\alpha$, structured ensemble are therefore seen to
behave in essentially the same way as the simple ensemble
Eq.~(\ref{eq:toy}).  The, only difference is that the condition of
validity of Eqs.~(\ref{eq:F:qgg1}) and (\ref{eq:F:qll1}), that is
respectively
	\begin{eqnarray}
	& & q \gg (\alpha_k D)/v_k^2 \label{eq:cond1}\\
	\mbox{and} &\qquad& v_k^{-1} \ll q \ll (\alpha_k D)/v_k^2
	\; , \label{eq:cond2}
	\end{eqnarray}
must now be fulfilled for all $k$'s.
The transition between the different regimes
of the distributions may therefore be less sharp than for the
ensemble Eq.~(\ref{eq:toy}).

If the partial barriers structures were to become highly developed,
say to the point that the ensemble could meaningfully be described in
terms of band matrices, then obviously the issue of localization
should have to be considered.  In this case, the orders of magnitude
of the $\alpha_k$ might become comparable to those of the $v_k$,
and most of the splitting distribution might be actually in a
transition-like regime.
This is exactly the sort of problems we pointed out in our earlier
discussion of the physical situation.   However, as long as
the $\alpha_k$'s  are clearly larger than the $v_k$'s,
the transition from one regime to another should still take
place on a short scale as compared to the range spanned by the
distribution. Physically speaking, this condition amounts to
saying that classically forbidden processes are always much slower
than classically allowed ones. In that case, the form of the result
should not be noticeably affected.  For example,  as seen
is Fig~\ref{fig:MCqo},  the distribution resulting
from the ensemble described in Eq.~(4.18) still perfectly follows
the predicted form Eqs.~(\ref{eq:distri:a})-(\ref{eq:distri:c})
(note however that $\alpha$ is now a tunable parameter).

\section{Conclusion}
\label{sec:conc}

As a conclusion, we have provided in this paper an analytical study of the
splitting distributions generated by ensembles of random matrices
constructed in \cite{tom94} to model a tunneling process
in the chaos assisted regime.  The original ensembles may contain such
a complicated structure that a general answer to this problem may
seem a priori out of reach.  Nevertheless, it turns out that
only the average size of the tunneling matrix elements and the
degree of correlation between the chaotic spectra in the different
symmetry class affect the distribution, and that therefore the problem
can be reduced to a simpler formulation which is tractable.

The basic reason for the considerable simplifications encountered
was in essence already pointed out in \cite{tom94}. It is due to the fact
that for large splittings only the situation of near-resonance to
a given state of the chaotic sea is of relevance. To this obvious
remark, we only need add that for the case in which efficient barriers
are at work, the tunneling operates not through single states, but through
quasi-degenerate doublets of states of opposite parity. These are
of course less efficient in promoting tunneling, since the particle
requires a time of the order of the width of the doublet to reach the
symmetrical torus.
In either case, the behavior is determined by rather natural
probabilistic considerations. It turns out to be sufficient
to consider only the probability of one single eigenvalue being
near the tunneling state, so that correlations between eigenvalues
and the like could be safely ignored. Further, the very simplicity
of the physical picture given here results in it being
fairly robust to changes in minor details of the model. Thus it
does not appear necessary  that all states in the chaotic
sea should participate equally in the tunneling process, nor
 that the couplings should be uniform. In fact, the main limitations
of our result seem to be
 the ones related to localization phenomena. If the
structure of the barriers in phase space is sufficiently complicated,
it is possible that localization effects, associated to the presence
of a large number of partial barriers, become as effective in limiting
tunneling from one quasimode to its symmetric partner as the initial
classically forbidden process.  In that case, the splitting
distributions we have obtained would not be relevant anymore.
However, this should not be a too severe limitation, and it should generally
be possible to determine for any given system whether this
takes place or not. When it does not, the picture of
tunneling in the presence as well as in the absence of barriers to
transport is indeed the one we gave.  This
is  substantiated by the numerical work
done: In particular, we showed that not only the simplest model of
a barrier gives results in good agreement with theoretical predictions,
but also a highly specific random matrix ensemble constructed
explicitly in order to model chaos-assited tunneling in a system of
coupled quartic oscillators was well fitted by the theoretical
predictions, as were also the actual splitting distribution for this
system.

This might possibly open up a way to identify chaos-assisted
tunneling in experimental systems. In such systems, the exhaustive study
of the classical mechanics necessary to produce a satisfactory
 random matrix
ensemble would probably not be feasible. Nevertheless the above remarks
strongly suggest that if chaos-assisted tunneling is present, the
splitting distribution will reflect the fact by showing a highly specific
and well-characterized behavior.  Indeed, as discussed throughout
this paper, only the scale of the distribution and the position of
the transitions between the different regimes are system dependent, but
the shape of the distribution is essentially universal.
In particular the experimental detection of a transition
from a  $\delta^{-3/2}$ behavior, characteristic of the $G$-like regime,
to a $\delta^{-2}$ behavior characteristic of the Cauchy-like regime,
 would  be a powerful argument in favor of the presence
of chaos-assisted tunneling.

\section*{Acknowledgments}

We acknowledge many helpful discussions with Oriol Bohigas,
as well as with Steven Tomsovic which we thank in addition for providing
us with the numerical data of the quartic oscillators system
used in Fig.~\ref{fig:qo}.  Denis Ullmo wants to thank  the
Institute for Nuclear Theory, Seattle WA, where parts of this work has
been done. Fran\c cois Leyvraz would like to thank the Division
de Physique Th\'eorique for its kind hospitality during
his sabbtical stay there where most of this work was done, as well as
acknowledge financial support by DGAPA during that same period.
The Division  de Physique Th\'eorique is ``Unit\'e de
Recherche des Universit\'es Paris~11 et Paris 6 Associ\'ee au CNRS''.

\appendix

\section{Computation of the Distribution Function}

Denote by brackets the integration over $e^+_i$ and $e^-_i$
with the weight function $P(e^+_i,e^-_i)$ (see Eq~(\ref{eq:joint}).
We define
	\begin{eqnarray}
	p(X) & = &
	\left\langle \delta \left(x- \bar v_t^2 \sum_{i=1}^N
	   \left(\frac{1}{e^+_i} - \frac{1}{e^-_i}\right)
	    \right) \right\rangle \; , \\
	F(q) & = & \int_{-\infty}^\infty F(X)e^{iqX}dx \nonumber \\
	& = & \left\langle \exp \left( {iq \bar v_t^2} \sum_{i=1}^N
	 \left(\frac{1}{e^+_i} - \frac{1}{e^-_i}\right)
		\right) \right\rangle
	\end{eqnarray}
This last expression factorizes in $N$ factors, each of which is a
double integral. Denoting the corresponding average over $e^+$
and $e^-$ also by brackets, one obtains:
	\begin{eqnarray}
	F(q) & = &
	\left\langle \exp \left( {iq \bar v_t^2} (1/e^+-1/e^-) \right)
	  \right\rangle^N \\
	& = & \left(1-
	\left\langle 1 -\exp\left({iq \bar v_t^2} (1/e^+-1/e^-) \right)
	      \right\rangle \right)^N  \nonumber \\
	& = & \left(1-\frac{I_N(q)}{N}\right)^N \nonumber
	\end{eqnarray}
where the last line defines $I_N(q)$. The reason for this manipulation
is that in this way $I_N$ goes to a finite limit $I$ as
$N \rightarrow \infty$ and therefore
	\begin{equation} \label{eq:eq_app}
	\lim_{N \to \infty} F(q) = \lim_{N \to \infty} (1-I(q)/N)^N
	= \exp(-I(q)) \; .
	\end{equation}
Note moreover that taking into account the fact that the tunneling matrix
elements are random variable of variance $v_t^2$ instead of being constant
just amount to understand $\langle \cdot \rangle$ as containing a further
integral on the tunneling matrix elements distribution.  This introduce
no further difficulties in the calculation of $I$, except for still
heavier notations.  We shall therefore not consider it in this appendix,
and just modify the final result in the  appropriate way at the end
of section~\ref{sec:DerOfEqs}.

 One finds
	\begin{equation}
	I(q)  =  {1\over\sqrt{2\pi} \bar \alpha}
	\int de^+\,de^-\,
	   \left(1 - \exp\left({i\bar v_t^2 q}(1/e^+ - 1/e^-)\right)\right)
	   \exp\left(-\frac{(e^+ - e^-)^2}{2 \bar \alpha^2}\right) \; ,
	\end{equation}
since, the above integral being convergent, $\lim_{N \to \infty} I_N$
is just obtained by dropping the term $-\pi(e^+/N)^2$ in the exponent
of $P(e^+,e^-)$ (see Eq.~(\ref{eq:joint})).  Making the successive
transformations $y = (1/e^+ + 1/e^-)/(1/e^+ - 1/e^-)$,
$w =   (1/e^+ - 1/e^-)$, followed by
$t=w (\bar \alpha (1-u^2)/\sqrt 8 )$, $I(q)$ can be expressed as
	\begin{eqnarray}
	   I(q)  & = & {8 \over \sqrt{2\pi} \bar \alpha}
	\int{dy \over (1-y^2)^2}
	   \int {dw \over \vert w\vert^3}(e^{iw  \bar v_t^2 q}-1)\
	   \exp\left(-{8 \over \bar \alpha^2 (1-y^2)^2 w^2}\right)
	   \nonumber \\
	& = & \frac{\bar \alpha}{\sqrt{2\pi}}
	 \int dy \int {dt \over \vert t \vert^3}
	     \left( 1-
	\cos \left( \frac{\sqrt 8 q \bar v_t^2}{\bar \alpha (1-y^2)}
	 \right)
              \right) e^{-1/t^2} \; .
	\end{eqnarray}
If we now introduce $\Phi(z)$ as in the text,
	\begin{equation}
	\Phi(z)  =  2\int_0^\infty{dt\over t^3}(1-\cos zt)e^{-1/t^2} \; ,
	\end{equation}
one easily obtains
	\begin{equation}
	I(q) = {\bar \alpha \over \sqrt{2\pi}} \int dy \,
	\Phi\left({\sqrt8 \bar v_t^2 q \over \bar \alpha \vert 1-y^2 \vert}
	      \right) \; .
	\end{equation}
 From this follows the formula given in the text:
	\begin{equation}
	F(q)  =
	    \exp\left(-{\bar \alpha \over\sqrt{2 \pi} }
	    \Psi(\bar v_t^2 q / \bar \alpha) \right)
	    \; ,
	\end{equation}
where $\Psi(\tilde q)$ is given by the expression
	\begin{equation}
	\Psi(\tilde q)  =  \int_{-\infty}^\infty dy\,
	   \Phi\left({\sqrt{8} \tilde q\over|1-y^2|}\right) \; .
	\end{equation}

\section{Some Useful Integrals}

We first give another expression for $\Phi(y)$:
	\begin{equation}
	\Phi(y)=\int_0^\infty dt\,(1-\cos{y\over\sqrt t})e^{-t},
	\end{equation}
which is obtained from the original definition by substituting
$1/t^2$ by $t$.{ From} this follows
	\begin{eqnarray}
	\int_{-\infty}^\infty\Phi(1/y)dy
	&=&\int_{-\infty}^\infty\Phi(y){dy\over y^2}\nonumber\\
	 &=&\int_0^\infty dt\,e^{-t}
	       \int_{-\infty}^\infty(1-\cos{y\over\sqrt t})
	        {dy\over y^2}\nonumber\\
	 &=&\int_0^\infty{dt\over\sqrt t}e^{-t}\int_{-\infty}^\infty
	 {1-\cos y\over y^2}dy\\
 	&=&\Gamma({1/2})\int_{-\infty}^\infty{\sin y\over y}dy=\pi^{3/2}
	 \nonumber
	\end{eqnarray}
The other integral is handled similarly:
	\begin{eqnarray}
	\int_{-\infty}^\infty\Phi(1/y^2)dy
	&=&\int_{-\infty}^\infty\Phi(y^2){dy\over y^2}\nonumber\\
	 &=&\int_0^\infty{dt\over t^{1/4}}e^{-t}\int_{-\infty}^\infty
	 {1-\cos y^2\over y^2}dy\\
	 &=&2\Gamma({3/4})\int_{-\infty}^\infty \sin y^2\,dy=\sqrt{2\pi}
	 \Gamma({3/4}).
	 \nonumber
	\end{eqnarray}

\newpage

\begin{figure}
\caption{Comparison between the quartic oscillator's tunneling splitting
distribution (square symbols) and the predicted form
Eqs.(\protect{\ref{eq:distri:a}})-(\protect{\ref{eq:distri:c}})
for two different groups of tunneling tori.  Except for their
presentation, the quartic oscillator's data are the same as those in
Fig.~13 of Ref.~\protect{\cite{tom94}}.
The transition from the $G$-like behavior (solid line) to Cauchy-like
behavior (chained dot) is clearly seen, in spite of this latter being
valid on a much shorter range.
  (a) Group $ T_0$
(using the notations of Ref.~\protect{\cite{tom94}}), with
 $v_t = 1.1\, 10^{-2}$.
   (b) Group $T_1$ , with $v_t = 2.5\, 10^{-2}$.  The transport parameter
  has the same value $\alpha=.04$ in both cases, consistent with
  the fact that the partial barriers structure is the same in both
  cases.  It has been taken into account that only a fraction
  $D_{\rm eff} = 0.36 D$ of the states are actually participating to
  the tunneling process.}
\label{fig:qo}
\end{figure}

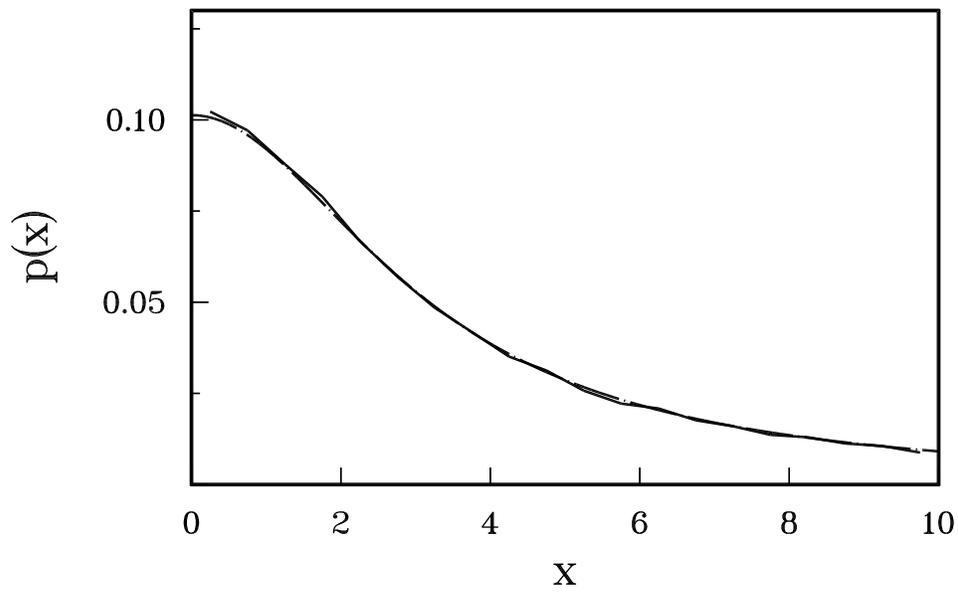
\begin{figure}
\caption{Comparison between a Monte Carlo calculated distribution
of the reduced variable $x=\delta^\pm (D/v_t^2)$
 (solid line) and the Cauchy
law Eq.~(\protect\ref{eq:cauchy1}) (chained dot).  The Monte Carlo
result is obtained from the numerical diagonalization of $10^5$
matrices of size $80 \times 80$, which matrix elements are taken
at random with the distribution specified by the ensemble
Eq.~(\protect\ref{eq:without}) (using $v_t/D=10^{-4}$).  It thus take
fully into account the GOE correlations of the chaotic states.
Nevertheless, and although a linear scale has been used to emphasize
the center of the distribution where the effects of correlations
should be the strongest, the two curves are essentially
undistinguishable.}
\label{fig:without}
\end{figure}

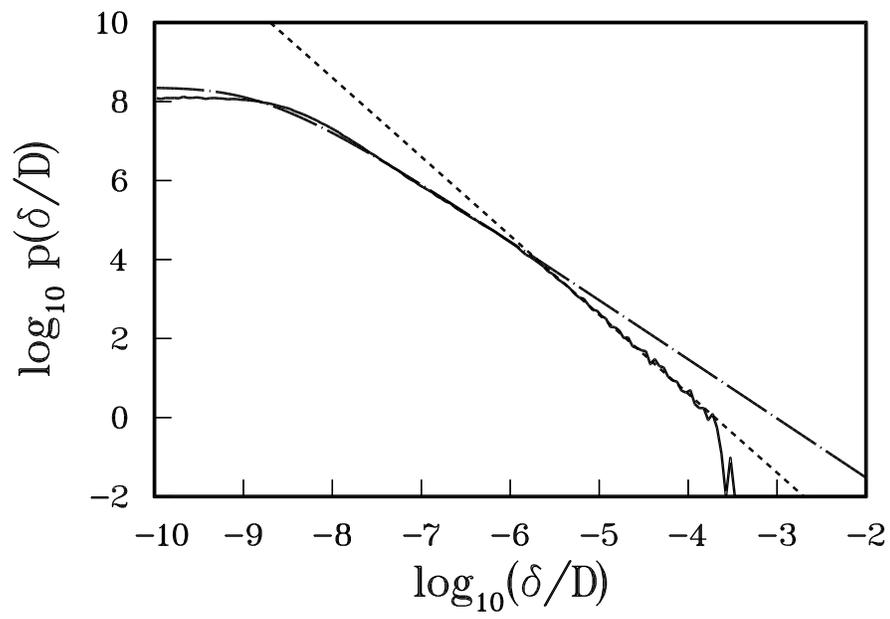
\begin{figure}
\caption{Comparison between a Monte Carlo calculated distribution
of splittings $\delta$ for the simple ensemble  Eq.~(\protect\ref{eq:toy})
(solid line) and the the predicted form
Eqs.(\protect{\ref{eq:distri:a}})-(\protect{\ref{eq:distri:c}}).
The parameters of the Monte Carlo calculations are $\Lambda =
10^{-2}/8$ (imposing $\alpha/D=0.1$ for the theoretical curve),
$v_t/D = 10^{-4}$, number of matrices:  $3. 10^5$,  size  of
matrices $60 \times 60$.
The three regimes: $G$-like behavior (chained dot), Cauchy-like
behavior (dash), and truncation of the Cauchy law for splitting
greater than $v_t/D$ are clearly seen.}
\label{fig:MCtoy}
\end{figure}

\begin{figure}
\caption{Comparison between a Monte Carlo calculated distribution
of splittings $\delta$ for the ensemble  Eq.~(\protect\ref{eq:qo_struct})
with $f_1=0.5, \;  f_2=0.2, \; f_3=0.3$, and
$\Lambda_{12} = 0.14, \; \Lambda_{13}=0.11$
(solid line), and the the predicted form
Eqs.(\protect{\ref{eq:distri:a}})-(\protect{\ref{eq:distri:c}}).
The  Monte Carlo calculations has been performed with $10^5$ matrices
of  size $100 \times 100$, using as tunneling parameter
$ (v_1)^2  /D = 10^{-3}$.
For the theoretical curves, namely the $G$-like (chained dot) and
Cauchy-like (dash) behaviors, the tunneling parameter is determined
by Eq.~(\protect\ref{eq:equi_v}) as
$(v_t)^2 = (v_1)^2 /2$.  The transport parameters is however here a tunable
parameter, which has been taken equal to $\alpha/D = 0.1$.}
\label{fig:MCqo}
\end{figure}

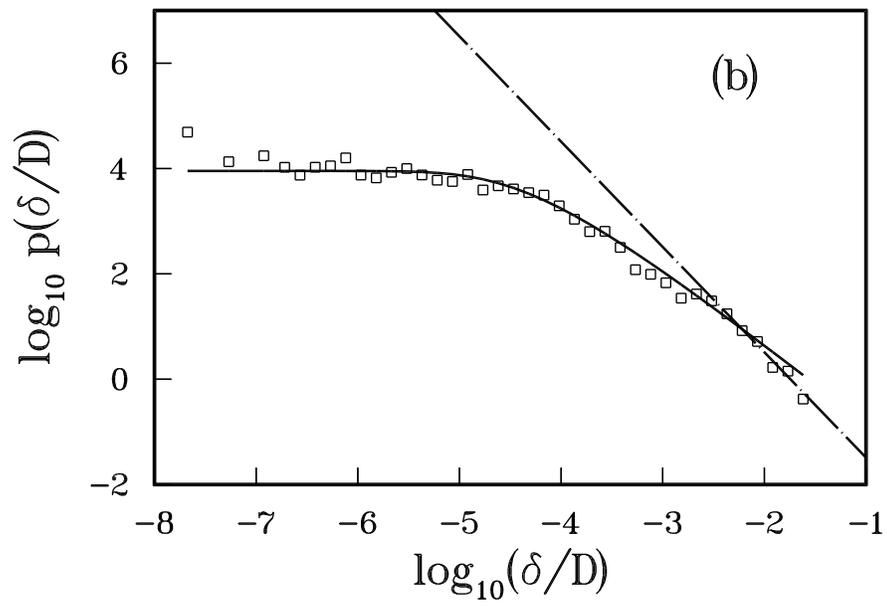
\end{document}